\documentstyle[prl,aps,multicol]{revtex}

\begin{document}
\draft
\preprint{MA/UC3M/02/1997}
\title{Dimensional crossover of the fundamental-measure functional for
parallel hard cubes}
\author{Jos\'e A.\ Cuesta and Yuri Mart\'\i nez--Rat\'on}
\address{Grupo Interdisciplinar de Sistemas Complicados
(GISC), \\ Departamento de Matem\'aticas, Escuela Polit\'ecnica
Superior, Universidad Carlos III de Madrid, \\ c/ Butarque, 15, 28911 --
Legan\'es, Madrid, Spain}

\maketitle

\begin{abstract}
We present a regularization of the recently proposed fundamental-measure
functional for a mixture of parallel hard cubes. The regularized
functional is shown to have right dimensional crossovers to any smaller
dimension, thus allowing to use it to study highly inhomogeneous phases
(such as the solid phase). Furthermore, it is shown how the functional
of the slightly more general model of parallel hard parallelepipeds can
be obtained using the zero-dimensional functional as a generating
functional. The multicomponent version of the latter system is also given,
and it is suggested how to reformulate it as a restricted-orientation model
for liquid crystals. Finally, the method is further extended to build a
functional for a mixture of parallel hard cylinders.
\end{abstract}

\begin{multicols}{2}
\narrowtext

Although nowadays the success of {\em density functional} theories to
describe nonuniform fluids and their phase transitions is out of
question, it is generally acknowledged that these theories are based on
much to empiric assumptions to be considered a systematic tool to study
liquids\cite{DFT}. Nearly all of the available functionals propose a
dependence on the density through nonlocal weights, whose choice rely
upon approximations around the uniform fluid combined with heuristic
arguments. Accordingly, they always need a good knowledge of the bulk
structure and thermodynamics of the uniform fluid, which thus becomes
an input of the functional. There is, however, a kind of functionals
which do not follow this general framework and are constructed only
from geometric features of the particles; these are the so called
{\em fundamental measure functionals} (FMFs)
\cite{rosenfeld1,kierlik,cuesta}. Two important features distinguish
FMFs from ``classical'' functionals: (1) the thermodynamics and
structure
of the uniform fluid (and even that of the nonuniform fluid) can be
{\em derived} from them, instead of requiring it as an input, and (2)
they achieve a dimensional reduction of the system by using strongly
inhomogeneous density profiles\cite{rosenfeld2,tarazona}. Actually
these two properties are interlaced, as it will become clear in what
follows.

The first FMF was proposed for a fluid of hard spheres (HS)
\cite{rosenfeld1,kierlik}, and it was obtained by assuming it to depend
on a small set of geometrically-weighted densities (how many and which
weights is dictated by the low density limit of the free-energy
functional). The precise functional dependence still required an ad
hoc scaled-particle assumption, but in turn, it yielded the
Percus-Yevick (PY) free energy and direct correlation function (DCF),
and allowed to compute further structural properties (such as the
third order DCF \cite{rosenfeld1}). The functional admits
two-dimensional (2D) \cite{rosenfeld3} as well as a one-dimensional
(1D) \cite{rosenfeld1} extensions, the latter being the well-known
{\em exact} functional.

There is a feature the exact HS functional exhibits: a right
dimensional crossover, or, in other words, the fact that the functional
for $D-1$ dimensions
must come out as a result of the one for $D$ dimensions when the
density profile is a delta function along one of the coordinates.
But this turns out to be a much to stringent requirement to be fulfilled
by an approximate functional. Needless to say that functionals with this
property will provide good descriptions of strongly inhomogeneous
systems (e.g.\ fluids confined by walls). However, it would not be of
great importance to describe bulk phases were it not because there is a
special dimensional crossover any functional willing to exhibit a
sensible
solid phase should verify: the reduction to a 0-dimensional (0D) system
\cite{rosenfeld2,rosenfeld3}, i.e.\ a cavity able to contain at most
one particle (which clearly mimics the situation of a particle in a
crystal lattice). None of the ``classical'' functionals verify any
dimensional crossover \cite{nota}, but the FMF for HS does some of them
\cite{rosenfeld2};
however, when trying to reduce from 3D to 0D, some divergences appear
that cause this functional not to be able to stabilize a solid phase.
Although an heuristic modification of the functional eliminates those
unphysical divergencies \cite{rosenfeld2}, it is very remarkable that
the D-dimensional HS FMF, for any dimension, can be obtained from the
exact 0D functional by requiring it to have a right dimensional
crossover from $D$ to
quasi-0 dimensions \cite{tarazona} (according to Ref.\ \onlinecite{tarazona},
a quasi-0D system is provided by a density function consisting of a sum
of deltas placed at points such that HS simultaneously placed at them
overlap). The results turn out to be non-local
functionals of the one-particle weighted densities, which can be
further approximated by local ones without loosing their main
dimensional crossovers. The moral of this is that the 0D functional
seems to be the only thing one needs in order to make a FMF for a given
system.

Recently, a FMF has been developed for a system of parallel hard cubes
(PHC) \cite{cuesta}. Although such a system is rather unrealistic, it
is far more suitable than the HS system to study the demixing transition
of mixtures \cite{cuesta,dijkstra}, and so, it deserves consideration.
On the other hand, on a fundamental viewpoint, having FMFs for systems
other than HS may help to clarify the basic structure of such
functionals in the aim of extending them to more general particle
shapes (arbitrarily orientable, anisotropic particles, for instance, so
as to study liquid-crystalline phases). In this letter we will show how
the functional proposed in \onlinecite{cuesta} can be regularized to
have right dimensional crossovers, and this will make clear a simple
relation between the
FMF for PHC in different dimensions which will permit to derive them
very easily from the 0D functional.

Any FMF expresses the excess (over the ideal) free energy of the system
in terms of the local density $\rho({\bf r})$ as
\begin{equation}
F^{\mbox{\scriptsize ex}}[\rho]=k_BT\int d{\bf r}\,
\Phi^{(D)}\bigl(\{n_{\alpha}({\bf r})\}\bigr),
\label{eq:0}
\end{equation}
where $D$ is the dimension and the $n_{\alpha}({\bf r})$ are a set of
suitably chosen geometrically-weighted densities. According to Ref.\
\onlinecite{cuesta}, the free energy density $\Phi^{(3)}$, for the
system of (3D) PHC, is given by $\Phi^{(3)}=\Phi^{(3)}_1+\Phi^{(3)}_2+
\Phi^{(3)}_3$, with
\begin{equation}
\begin{array}{rclrcl}
\Phi^{(3)}_1 & = & -n_0\ln(1-n_3), &
\Phi^{(3)}_2 & = & \displaystyle
    \frac{{\bf n}_1\cdot{\bf n}_2}{1-n_3}, \\
\Phi^{(3)}_3 & = & \displaystyle
    \frac{5n_2^3-9n_2{\bf n}_2\cdot{\bf n}_2}{54(1-n_3)^2}. & & &
\end{array}
\label{Phi123}
\end{equation}
Here $n_{\alpha}\equiv\rho\otimes
\omega_{\alpha}$ and ${\bf n}_{\alpha}\equiv\rho\otimes
{\bf w}_{\alpha}$, where $\otimes$ denotes a convolution. The weights
$\omega_{\alpha}$ and ${\bf w}_{\alpha}$ are given by
\begin{equation}\begin{array}{rclrcl}
\omega_0({\bf r}) & = & \zeta_x\zeta_y\zeta_z, &
\omega_3({\bf r}) & = & \tau_x\tau_y\tau_z, \\
{\bf w}_1({\bf r}) & = & \multicolumn{4}{l}{(\tau_x\zeta_y\zeta_z,
     \zeta_x\tau_y\zeta_z,\zeta_x\zeta_y\tau_z),} \\
{\bf w}_2({\bf r}) & = & \multicolumn{4}{l}{(\zeta_x\tau_y\tau_z,
     \tau_x\zeta_y\tau_z,\tau_x\tau_y\zeta_z),} \\
\omega_1({\bf r}) & = & {\bf u}\cdot{\bf w}_1({\bf r}), &
\omega_2({\bf r}) & = & {\bf u}\cdot{\bf w}_2({\bf r}),
\end{array}
\label{w-3D}
\end{equation}
with $\tau_x=\Theta(\sigma/2-|x|)$, $\zeta_x=(1/2)\delta(\sigma/2-|x|)$
($\sigma$ being the edge-length of the cubes), and ${\bf u}=(1,1,1)$.
Notice the different normalization in the definition of the weighted
densities with respect to that of Ref.\ \onlinecite{cuesta}.

As mentioned in \onlinecite{cuesta}, the 2D functional can be obtained
in a similar way; in this case the weights are given by
\begin{equation}
\begin{array}{rclrcl}
\omega_0({\bf r}) & = & \zeta_x\zeta_y, &
\omega_2({\bf r}) & = & \tau_x\tau_y,  \\
{\bf w}_1({\bf r}) & = & (\tau_x\zeta_y,\zeta_x\tau_y), &
\omega_1({\bf r}) & = & {\bf u}\cdot{\bf w}_1({\bf r}),
\end{array}
\label{w-2D}
\end{equation}
with ${\bf u}=(1,1)$, and $\Phi^{(2)}=\Phi^{(2)}_1+\Phi^{(2)}_2$, with
\begin{equation}
\Phi^{(2)}_1=-n_0\ln(1-n_2), \quad
\Phi^{(2)}_2=\frac{n_1^2-{\bf n}_1\cdot{\bf n}_1}{2(1-n_2)}.
\label{Phi12}
\end{equation}

Of course, the 1D functional is the same as that of HS (i.e.\ the exact
one) because in this case both, HS and PHC, reduce to segments on a line.
Therefore, $\Phi^{(1)}=\Phi^{(1)}_1=-n_0\ln(1-n_1)$, where
$\omega_0({\bf r})=\zeta_x$ and $\omega_1({\bf r})=\tau_x$.

Let us now consider the 1D to 0D dimensional crossover. To this purpose
let us define
the class of quasi-0D densities as the set of functions with support
in the interval $(-\sigma/2,\sigma/2)$ (the fact that this interval is
centered around zero is completely irrelevant and it is only assumed
for simplicity). The densities $\rho(x)=\delta(x)$ and
$\rho(x)=\sum_i\eta_i\delta(x-x_i)$, with $-\sigma/2<x_i<\sigma/2$,
$\eta_i\geq 0$, and $\sum_i\eta_i=\eta\leq 1$ (those considered in
\cite{tarazona}), belong to this class, but smooth functions with
support within that interval do it as well. Then it is straightforward
to check that \cite{nota2}
\begin{equation}
n_0=-\frac{1}{2}s_x\frac{\partial n_1}{\partial x},
\label{n0-n1'}
\end{equation}
where $s_x=\mbox{sign}\,x$. Now, if we consider that
\begin{equation}
-\ln(1-n_1)\frac{\partial n_1}{\partial x}=
\frac{\partial\Phi^{(0)}(n_1)}{\partial x},
\label{Phi0'}
\end{equation}
where \cite{rosenfeld2} $\Phi^{(0)}(\eta)=\eta+(1-\eta)\ln(1-\eta)$, we
can check that $F^{\mbox{\scriptsize ex}}=\int dx\,\Phi^{(1)}=
\Phi^{(0)}(n_1(0))$,
which reduces to the excess free energy of the 0D system for {\em any}
density of the quasi-0D class.

To check the 2D to 1D and 2D to 0D dimensional crossovers, let us
define the quasi-1D
density class as the set of 2D functions with support in the
band $-\sigma/2<y<\sigma/2$, and let us redefine the quasi-0D density
class as the subset of the quasi-1D class whose members vanish for
$x\notin(-\sigma/2,\sigma/2)$ (i.e.\ 2D functions with support in a
square centered at the origin). Let us consider now the 2D functional
(\ref{Phi12}), and let us put ${\bf n}_1=(n_{1,x},n_{1,y})$; then
$n_1^2-{\bf n}_1\cdot{\bf n}_1=2n_{1,x}n_{1,y}$. On the other hand, as
in the previous case,
\begin{equation}
n_0=-\frac{1}{2}s_y\frac{\partial n_{1,x}}{\partial y},  \quad
n_{1,y}=-\frac{1}{2}s_y\frac{\partial n_2}{\partial y};
\label{n0-n1'n1'-n2}
\end{equation}
therefore
\begin{equation}
\Phi^{(2)}=-\frac{1}{2}s_y\frac{\partial}{\partial y}
\Bigl[-n_{1,x}\ln(1-n_2)\Bigr],
\label{Phi2-Phi1'}
\end{equation}
and then
\begin{eqnarray}
F^{\mbox{\scriptsize ex}} & = & \int dxdy\,\Phi^{(2)}   \nonumber  \\
 & = & \int dx\,\Phi^{(1)}(n_{1,x}(x,0),n_2(x,0)),
\label{FexPhi1}
\end{eqnarray}
which is the right 2D to 1D dimensional crossover. Furthermore, as we
have discussed previously, the functional (\ref{FexPhi1}) has the right
dimensional crossover to 0D for any quasi-0D density.

In both cases, 1D and 2D, the right dimensional crossover arises as a
consequence of
the fact that for quasi-0D densities the functionals can be obtained as
\begin{eqnarray}
\Phi^{(1)} & = & -\frac{1}{2}s_x
     \frac{\partial\Phi^{(0)}}{\partial x},       \label{Phi0x} \\
\Phi^{(2)} & = & \left(-\frac{1}{2}\right)^2s_{xy}
     \frac{\partial^2\Phi^{(0)}}{\partial y\partial x}.  \label{Phi0xy}
\end{eqnarray}
The dimensional crossovers then follow from the direct integration of
the derivatives.
It can then be inferred that the D-dimensional functional with the
right dimensional crossovers will, for quasi-0D densities, be expressed as
\begin{equation}
\Phi^{(D)}=\left(-\frac{1}{2}\right)^Ds_{x_1\cdots x_D}
     \frac{\partial^D\Phi^{(0)}}{\partial x_D\cdots\partial x_1};
\label{Phi0x1...xD}
\end{equation}
but if we compute the 3D case, and take into account that
\begin{equation}
\begin{array}{rcl}
{\bf n}_2 & = & \displaystyle \left(-s_x
    \frac{\partial n_3}{\partial x},-s_y
    \frac{\partial n_3}{\partial y},-s_z
    \frac{\partial n_3}{\partial z}\right), \vspace*{2mm} \\
{\bf n}_1 & = & \displaystyle \left(s_{yz}
    \frac{\partial^2 n_3}{\partial y\partial z},s_{zx}
    \frac{\partial^2 n_3}{\partial z\partial x},s_{xy}
    \frac{\partial^2 n_3}{\partial x\partial y}\right),
    \vspace*{2mm} \\
n_0 & = & \displaystyle -s_{xyz}
    \frac{\partial^3 n_3}{\partial x\partial y\partial z},
\end{array}
\label{n2n1n0}
\end{equation}
we find that $\Phi^{(3)}_1$ and $\Phi^{(3)}_2$ are the same as those
in Eq.\ (\ref{Phi123}), but $\Phi^{(3)}_3$ is given by
\begin{equation}
\Phi^{(3)}_3=\frac{n_2^3-3n_2{\bf n}_2\cdot{\bf n}_2-
    2{\bf n}_2\cdot{\bf n}_2\cdot{\bf n}_2}{6(1-n_3)^2},
\label{newPhi3}
\end{equation}
where the notation ${\bf v}\cdot{\bf v}\cdot{\bf v}=v_x^3+v_y^3+v_z^3$
has been introduced. This new expression differs from that of
Eq.\ (\ref{Phi123}) not only in the numerical coefficients, but also
in the appearance of a new term, ${\bf n}_2\cdot{\bf n}_2\cdot
{\bf n}_2$, which has no equivalent in HS \cite{rosenfeld1,tarazona}.
At first this may seem a spurious term, because it is not even
rotationally invariant; on second thoughts, there is no reason why it
should be, since the system itself (PHC) is not rotationally invariant
either. Such a symmetry is replaced by a discrete group of symmetries
which, restricted to the fact that the vector densities ${\bf n}_1$
and ${\bf n}_2$ must have positive components, translates into a
simple axes-exchange symmetry. The new term obviously has this
symmetry. The important thing is that while this new 3D functional has
been {\em built} to have right dimensional crossovers, the old one
does not even fulfill the 3D to 0D one ($\Phi^{(3)}_3$ develops delta
cube divergences when the density approaches a delta function, much in
the same way as the old version of the HS FMF did \cite{rosenfeld1}).
It can be checked that the new functional stabilizes a solid phase, for
packing fractions above $\eta\approx 0.3$, by following the standard
procedure of parametrizing the density function by a sum of gaussians
placed at the lattice sites \cite{DFT}, whereas the divergences in
the old functional give rise to a free energy which decreases monotonously
as the gaussians shrink.

But to be correct, the new 3D functional should provide the exact second
and third virial coefficients in the density expansion of the DCF, for
this is the requirement under which FMF are built
\cite{rosenfeld1,cuesta}. It is very remarkable that, when applied to
the uniform fluid, the new functional provides {\em exactly the same
expressions} for both the free energy and the DCF as found in Ref.\
\onlinecite{cuesta} (bear in mind the different normalization of the
weighted densities!). Therefore, the new term introduces nothing new for
the uniform fluid, but it drastically changes the behavior of the
nonuniform fluid.

Formula (\ref{Phi0x1...xD}) provides a simple method to obtain the FMF
for a D-dimensional PHC fluid; although the formula only holds true for
quasi-0D densities, once the functional form has been derived and the
scalar, vector, and tensor weighted densities have been identified, it
can also be applied to arbitrary density functions [this is precisely
the procedure we have followed to obtain the 3D functional
(\ref{newPhi3})]. At this level, Eq.\ (\ref{Phi0x1...xD}) may seem
nothing more than a fortunate trick to obtain the functionals; however,
a slight manipulation of the equation will provide further insight on
the actual meaning of this method.

First of all, let us consider the slightly more general problem of a fluid
of D-dimensional parallel hard parallelepipeds (PHP) of edge-lengths
$\sigma^{(1)},\dots,\sigma^{(D)}$ along the $x_1,\dots,x_D$ axes. If we
define $\eta({\bf r})=\rho\otimes\omega_D({\bf r})$, with
\begin{equation}
\omega_D({\bf r})=
\Theta\left(\frac{\sigma^{(1)}}{2}-|x_1|\right)\cdots
\Theta\left(\frac{\sigma^{(D)}}{2}-|x_D|\right),
\label{eta(r)}
\end{equation}
then, the 0D free-energy density functional for this system will be
given by
\begin{equation}
\Phi^{(0)}_{\mbox{\scriptsize PHP}}=\eta({\bf r})+
[1-\eta({\bf r})]\ln[1-\eta({\bf r})].
\label{Phi0DPHP}
\end{equation}
Formula (\ref{Phi0x1...xD}) still yields the D-dimensional functional
for quasi-0D densities, but if we make use of the identity
\[
-\frac{1}{2}s_{x_j}\frac{\partial}{\partial x_j}
\Theta\left(\frac{\sigma^{(j)}}{2}-|x_j|\right)=
\frac{\partial}{\partial\sigma^{(j)}}
\Theta\left(\frac{\sigma^{(j)}}{2}-|x_j|\right),
\]
Eq.\ (\ref{Phi0x1...xD}) transforms into
\begin{equation}
\Phi^{(D)}_{\mbox{\scriptsize PHP}}=
\frac{\partial^D\Phi^{(0)}_{\mbox{\scriptsize PHP}}}
{\partial\sigma^{(D)}\cdots\partial\sigma^{(1)}},
\label{Phi0s1...sD}
\end{equation}
which is valid not only for quasi-0D densities, but for {\em completely
general} density functions (because it does not involve derivatives with
respect to the coordinates).

Equation (\ref{Phi0s1...sD}) has the important consequence that all the
thermodynamics of the inhomogeneous D-dimensional fluid is contained
in the 0D functional. This result reveals the same fundamental idea
recently formulated for HS \cite{tarazona}. In the latter, however, the
derivation is not as simple as for the PHP system. An attempt to find an
equivalent to Eq.\ (\ref{Phi0s1...sD}) for HS, where the 0D functional
acts as a generating functional, has just been carried out
\cite{gonzalez}. In this work, the D-dimensional FMF for HS, in the
version of Ref.\ \cite{kierlik}, in which all weighted functions are
scalar, is given by the action on the 0D functional of a differential
operator with respect to the sphere radius. The result is nevertheless
not as satisfactory as for the PHP system (for instance, the 3D DCF
generates an unphysical delta function at contact). On the other hand,
while Eq.\ (\ref{Phi0s1...sD}) emerges here in a very natural way, the
differential operator in \cite{gonzalez} is introduced heuristically
with some free constants which are fitted afterwards by imposing certain
constraints (e.g.\ to match a given equation of state). On the contrary,
the functional for 2D agrees with the simplified version of the one
obtained in \cite{tarazona}, so that the method seems promising and, in
view of what happens for PHP, it deserves further consideration.

Generalizing (\ref{Phi0s1...sD}) to mixtures is straightforward; it
simply amounts to writing
\begin{equation}
\Phi^{(D)}_{\mbox{\scriptsize PHP}}=\sum_i
\frac{\partial^D\Phi^{(0)}_{\mbox{\scriptsize PHP}}}
{\partial\sigma^{(D)}_i\cdots\partial\sigma^{(1)}_i},
\label{mix}
\end{equation}
where $\sigma^{(j)}_i$ denotes the edge-length along the $x_j$
coordinate of species $i$. Of course, now $\eta({\bf r})=\sum_i
\rho_i\otimes\omega^{(D)}_i({\bf r})$, where $\omega^{(D)}_i$ is given
by Eq.\ (\ref{eta(r)}) with the $\sigma^{(j)}$s corresponding to
species $i$. The functional (\ref{mix}) has a particularly important
application to study liquid-crystalline phases, for if we consider the
six possible orientations of a parallelepiped as belonging to six
different species, the functional (\ref{mix}) can be understood as
representing a restricted-orientation, one-species system of hard
parallelepipeds.  Besides that, the functional may provide some clues
to understand the structure of a FMF for freely-orientable, general
anisotropic particles.

The derivation we have made here suggests further extensions of the
theory. For instance, consider a system of hard parallel cylinders
oriented along the $z$ axis. It is clear that the 3D to 2D dimensional
crossover, eliminating the $z$ coordinate,
will transform this system into a system of hard disks. Therefore, if
$\Phi^{(2)}_{\mbox{\scriptsize HS}}$ is a FMF for hard disks (e.g.\
those of Refs.\ \cite{tarazona,gonzalez}), the corresponding FMF for
the cylinders can be derived from
$\widetilde\Phi^{(2)}_{\mbox{\scriptsize HS}}$, the same
functional as $\Phi^{(2)}_{\mbox{\scriptsize HS}}$, but with the weights
multiplied by $\Theta\left(L/2-|z|\right)$ ($L$ being the length of the
cylinders) and the 2D density of disks replaced by a 3D density of
cylinders. The derivation will simply be
\begin{equation}
\Phi^{(3)}_{\mbox{\scriptsize cyl}}=\frac{\partial
\widetilde\Phi^{(2)}_{\mbox{\scriptsize HS}}}{\partial L}.
\label{Phicyl}
\end{equation}
This functional can immediately be generalized to a mixture of parallel
hard cylinders as in (\ref{mix}), and this provides a tool to study
analytically the influence of polydispersity in the phase diagram of
liquid crystals \cite{stroobants}.

To summarize, we have shown how the FMF for PHC presented in
\cite{cuesta} can be regularized to have a right dimensional crossover
to any smaller
dimension. In the derivation we have found the remarkable result
that the FMF for PHP, in any dimension $D$, can be obtained by simply
deriving with respect to every edge-length the 0D functional, thus
prompting the idea of this functional being a kind of generating
functional (an idea already explored for HS \cite{tarazona,gonzalez},
but which for the present system reaches its clearest and simplest
expression). In passing we have obtained the FMF for a mixture of PHP,
and suggested how it can be exploited as a restricted-orientation model
for liquid crystals. Finally a further extension of the method permits to
find a FMF for a mixture of parallel hard cylinders, thus opening the
possibility of studying polydispersity in liquid crystals.

We are indebted to Y.\ Rosenfeld for calling our attention on this
problem; to P.\ Tarazona for illuminating discussions and for keeping
us informed of his progress in studying HS; to J.A.\ White for sending
us their preprint, and to B.\ Mulder for suggesting the extension
of the PHP functional to a restricted-orientation model. One of us
(J.A.C.) owes much to a discussion with L.\ Araujo which turned out to
be crucial for the development of this work.

\end{multicols}


\begin{thebibliography}{88}
\bibitem{DFT} See, for instance, the reviews M.\ Baus, J.\ Phys.:
    Cond.\ Matter {\bf 2}, 2241 (1990); {\em Fundamentals of
    Inhomogeneous Fluids}, D.\ Henderson ed.\ (Dekker, New York, 1992);
    G.\ J.\ Vroege and H.\ N.\ W.\ Lekkerkerker, Rep.\ Prog.\ Phys.\
    {\bf 55}, 1241 (1992); H.\ L\"owen, Phys.\ Rep.\ {\bf 237}, 249
    (1994).
\bibitem{rosenfeld1} Y.\ Rosenfeld, \prl {\bf 63}, 980 (1989); see
    also Y.\ Rosenfeld, J.\ Phys.: Cond.\ Matter {\bf 8}, 9289 (1996),
    and references therein.
\bibitem{kierlik} E.\ Kierlik and M.\ L.\ Rosinberg, \pra {\bf 42},
    3382 (1990); \pra {\bf 44}, 5025 (1991).
\bibitem{cuesta} J.\ A.\ Cuesta, \prl {\bf 76}, 3742 (1996).
\bibitem{rosenfeld2} Y.\ Rosenfeld, M.\ Schmidt, H.\ L\"owen, and P.\
    Tarazona, J.\ Phys.: Cond.\ Matter {\bf 8}, L577 (1996); preprint
    (1996).
\bibitem{tarazona} P.\ Tarazona and Y.\ Rosenfeld, preprint (1996).
\bibitem{rosenfeld3} Y.\ Rosenfeld, \pra {\bf 42}, 5978 (1990).
\bibitem{rosenfeld4} Y.\ Rosenfeld, J.\ Phys.: Cond.\ Matter {\bf 8},
    L795 (1996).
\bibitem{nota} And yet they describe accurately the solid phase
    \cite{DFT}. But this solid has unphysical properties, such as a
    negative density of vacancies, or a regular behavior of the free
    energy at closest packing \cite{rosenfeld2,rosenfeld4}.
\bibitem{dijkstra} M.\ Dijkstra and D.\ Frenkel, \prl {\bf 72}, 298
    (1994); M.\ Dijkstra, D.\ Frenkel and J.-P.\ Hansen, \jcp
    {\bf 101}, 3179 (1994).
\bibitem{nota2} If instead of considering this class of functions we
    take that of functions highly localized within the same interval,
    Eq.\ (\ref{n0-n1'}) becomes an asymptotic equality, so that the
    higher the localization the more accurate the equation.
\bibitem{gonzalez} A.\ Gonz\'alez, J.\ A.\ White, and R.\ Evans,
    preprint (1996).
\bibitem{stroobants} A.\ Stroobants, J.\ Phys.: Cond.\ Matter {\bf 6},
    A285 (1994).
\end{thebibliography}
\end{document}